\documentclass[showpacs,nofootinbib,preprintnumbers,prd,superscriptaddress,twocolumn]{revtex4-2}

\usepackage{amsmath}
\usepackage{braket}
\usepackage{amsfonts}
\usepackage{amssymb}
\usepackage{graphicx}
\usepackage[titletoc]{appendix}
\usepackage{color}
\usepackage{hyperref}
\usepackage{cleveref}
\usepackage[rightcaption]{sidecap}
\usepackage{subfigure}
\usepackage{comment}

\usepackage{float}
\usepackage{wasysym}
\usepackage{amssymb}

\usepackage{dcolumn}

\usepackage{array}
\usepackage{ctable}
\usepackage{multirow}
\usepackage{siunitx}
\usepackage{longtable}
\usepackage{tabularx}
\usepackage{booktabs}

\usepackage{tensor}
\usepackage{mathtools}
\usepackage{dirtytalk}
\usepackage{siunitx}
\DeclareSIUnit{\parsec}{pc}

\graphicspath{{Graphics/}}

\def\be{\begin{equation}}
\def\ee{\end{equation}}
\def\bea{\begin{eqnarray}}
\def\eea{\end{eqnarray}}
\renewcommand{\d}{\mathrm{d}}

\newcommand{\virgolette}[1]{``#1''}
\definecolor{vividviolet}{rgb}{0.62, 0.0, 1.0}
\definecolor{amaranth}{rgb}{0.9, 0.17, 0.31}
\definecolor{palatinateblue}{rgb}{0.15, 0.23, 0.89}
\definecolor{brightpink}{rgb}{1.0, 0.0, 0.5}
\definecolor{cornflowerblue}{rgb}{0.39, 0.58, 0.93}
\definecolor{deepcarminepink}{rgb}{0.94, 0.19, 0.22}
\definecolor{radicalred}{rgb}{1.0, 0.21, 0.37}

\hypersetup{ linktoc=all,
    colorlinks, linkcolor={palatinateblue},
    citecolor={brightpink}, urlcolor={amaranth}
}

\begin{document}

\title{Entanglement entropy in quantum black holes}

\author{Alessio Belfiglio}
\email{alessio.belfiglio@unicam.it}
\affiliation{Universit\`a di Camerino, Divisione di Fisica, Via Madonna delle Carceri, 62032, Italy.}
\affiliation{Istituto Nazionale di Fisica Nucleare (INFN), Sezione di Perugia, Perugia, 06123, Italy.}

\author{Orlando Luongo}
\email{orlando.luongo@unicam.it}
\affiliation{Universit\`a di Camerino, Divisione di Fisica, Via Madonna delle Carceri, 62032, Italy.}
\affiliation{Istituto Nazionale di Fisica Nucleare (INFN), Sezione di Perugia, Perugia, 06123, Italy.}
\affiliation{Department of Mathematics and Physics, SUNY Polytechnic Institute, Utica, NY 13502, USA.}
\affiliation{Istituto Nazionale di Astrofisica (INAF), Osservatorio Astronomico di Brera, Milano, Italy.}
\affiliation{Al-Farabi Kazakh National University, Almaty, 050040, Kazakhstan.}

\author{Stefano Mancini}
\email{stefano.mancini@unicam.it}
\affiliation{Universit\`a di Camerino, Divisione di Fisica, Via Madonna delle Carceri, 62032, Italy.}
\affiliation{Istituto Nazionale di Fisica Nucleare (INFN), Sezione di Perugia, Perugia, 06123, Italy.}

\author{Sebastiano Tomasi}
\email{sebastiano.tomasi@unicam.it}
\affiliation{Universit\`a di Camerino, Divisione di Fisica, Via Madonna delle Carceri, 62032, Italy.}
\affiliation{Istituto Nazionale di Fisica Nucleare (INFN), Sezione di Perugia, Perugia, 06123, Italy.}

\date{\today}

\begin{abstract}
We discuss the entanglement entropy for a massive Klein-Gordon field in two Schwarzschild-like quantum black hole spacetimes, also including a nonminimal coupling term with the background scalar curvature. To compute the entanglement entropy, we start from the standard spherical shell discretisation procedure, tracing over the degrees of freedom residing inside an imaginary surface. We estimate the free parameters for such quantum metrics through a simple physical argument based on Heisenberg uncertainty principle, along with alternative proposals as asymptotic safety, trace anomaly, and graviton corpuscular scaling. Our findings reveal a significant decrease in entropy compared to the area law near the origin for the quantum metrics. In both scenarios, the entanglement entropy converges to the expected area law sufficiently far from the origin. We then compare these results to the entropy scaling in regular Hayward and corrected-Hayward spacetimes to highlight the main differences with such regular approaches.
\end{abstract}



\maketitle
\tableofcontents

\section{Introduction}

In standard thermodynamics of classical systems, entropy is typically an extensive quantity. It is a measure of the amount of information we are missing about the system, preventing us from identifying the exact microstate that corresponds to the observed macrostate. This is not the case of quantum mechanics, where positive entropies may arise even when the state of the system is known with certainty. In fact, if the system can be divided into subsystems, and if the different subsystems are \emph{entangled}, then the entropy of a subsystem can be different from zero \cite{hor}.  The ground state \emph{von Neumann entropy} of a free scalar field in flat spacetime was initially computed in \cite{qft1}, showing its proportionality to the boundary area of the region under analysis. The same result was later obtained for a spherical entangling surface, deriving the reduced density operator for scalar degrees of freedom residing inside an imaginary sphere \cite{qft2}. The entropy defined in such calculations has become known as \emph{entanglement entropy}, as it can be used as an entanglement measure for pure states \cite{qit,pl1,Solodukhin:2011gn}. In contrast to thermal states, such entropy does not originate from our ignorance about the microstate of the system, but it is intrinsic and cannot be removed even at zero temperature \cite{entr1,entr2}. Surprisingly, the ground state entanglement entropy of many quantum systems is not an extensive quantity, but it rather scales as the area of the boundary of the subsystem under analysis, usually up to small logarithmic corrections. This peculiar scaling property of the entropy is now known as the \emph{area law} \cite{rev1}.

In recent years, the relevance of entanglement entropy in characterising quantum systems has attracted significant attention across various fields of physics, such as condensed matter physics \cite{condmat1, condmat2, condmat3, condmat4, condmat5,Brandão2013,Anshu2022}, quantum information processing \cite{ben1, ben2, ben3} and quantum field theory \cite{qft1, qft2, qft3, PasqualeCalabrese_2004,Casini_2009}. In particular, the emergence of an area law behaviour in discrete field theories shows striking similarities with the Bekenstein-Hawking entropy for black holes, which is proportional to the area of the black hole horizon \cite{bh1, bh2, bh3, bh4}, also suggesting possible connections with the holographic principle \cite{holo}. Accordingly, entanglement entropy has become a leading candidate in the attempt to understand the nature of black hole entropy and its quantum origin (see e.g. \cite{Das2010} for a review). More recently, seminal papers \cite{qft1,qft2} were indeed generalised to the case of static spherically symmetric configurations, confirming that entanglement entropy satisfies an area law when the scalar field is in its ground state \cite{Das:2007mj}. In this way, the scalar field entropy can be truly associated with the existence of an horizon, as in the case of black holes\footnote{The same result is obtained in \cite{qft1,qft2} by artificially creating an horizon in flat spacetime, see e.g. \cite{Das:2007mj} for a discussion on this point.}.

Since the ultimate nature of black hole entropy and, more broadly, black hole thermodynamics is necessarily quantum, there have been several attempts to construct quantum models of black holes in recent years (see e.g. \cite{Binetti:2022xdi} and references therein), with the aim of building effective frameworks to investigate quantum effects without resorting to a still unknown theory of quantum gravity. In this direction, one may for example introduce new physical quantities at Planck scales \cite{Binetti:2022xdi}, or move to the corpuscular picture, for which geometry should only emerge at suitable macroscopic scales from the underlying quantum field theory of gravitons \cite{casadio2022geometry}. Accordingly, it seems interesting to verify the existence of an area law for quantum fields within these effective frameworks and quantify possible deviations arising from quantum effects.

An evident issue that arises when trying to identify black hole entropy in terms of entanglement is related to the UV divergences appearing for the von Neumann entropy of quantum fields\footnote{See \cite{Solodukhin:2011gn} for a discussion on other possible issues associated with this interpretation and different approaches to solve them.}. Such divergences are typically removed by discretising field theories in a lattice, where the lattice spacing serves as the ultraviolet cutoff length \cite{qft1,qft2}.  When dealing with effective theories of quantum black holes, the Planck length emerges as the natural cutoff scale to describe possible deviations from the classical theory. The ground state of the discretised field is then computed; it is a pure state that can be described by the system's density operator $\rho$. Then, the presence of an horizon naturally divides the field degree of freedom into two distinct regions. Therefore, we trace the degrees of freedom inside the horizon, thus focusing only on the outside system $\rho_O=\mathrm{Tr}_I(\rho)$. The entanglement entropy $S=-\mathrm{Tr}(\rho_O\ln{\rho_O})$ is found to be proportional to $\mathcal{A}_H/a^2$, where $\mathcal{A}_H$ represents the area of the horizon, and $a$ is the lattice spacing.

Another relevant issue is related to the presence of nonminimal coupling, which may significantly affect entanglement generation for quantum fields. In different relativistic scenarios, such as early universe dynamics or regions near black holes \cite{faki, mae}, the effects of nonminimal field-curvature coupling seem to be relevant. They also predict novel phenomena, such as particle creation \cite{blm1,blm2}, non-equivalence between frames \cite{fara}, large-scale dynamical effects \cite{lm1}, vacuum energy cancellation \cite{lm2,lm3}, and so on. Within discrete scalar field theories, it has been proven that a coupling with the scalar curvature can lead to violations of the area law \cite{Belfiglio:2023sru}, so the same effect requires further investigation at Planck scales for effective approaches to quantum black holes.

Motivated by the above picture, in this work we aim to investigate the effects of a nonminimal coupling term on the entanglement entropy scaling for a massive Klein-Gordon field in quantum black hole spacetimes. We begin by estimating the radius at which quantum effects should become relevant in a Schwarzschild black hole scenario, in order to quantify the values of the free parameters of the quantum black hole models. In particular, we first focus on the metric proposed by \cite{Binetti:2022xdi}, which should result in a consistent way to characterise quantum corrections to the Schwarzschild metric, without committing to any particular quantum gravity model. We demonstrate that nonminimal coupling may lead to significant departures from area law at Planck scales. We then move to the corpuscular picture proposed in Ref. \cite{casadio2022geometry}, where Schwarzschild geometry is described in terms of the coherent state of a massless scalar field. Also in this case, we find significant violations to the area law close to the origin, due to the behaviour of the scalar curvature at these scales. Finally, we compare the entanglement entropy scaling in the above mentioned quantum metrics with Hayward-like regular black hole approaches.  We focus in particular on the Hayward metric \cite{PhysRevLett.96.031103} and a possible modification \cite{Knorr:2022kqp}, which is consistent with the principle of least action at large distances. We show that area law is preserved for the Hayward metric sufficiently close to the origin, due to the de Sitter core of this solution, while quantum gravity corrections may alter the entropy scaling at Planck scales, confirming that modifications occurring at Planckian distances from the would-be singularity can affect entanglement generation.

The paper is structured as follows. In Sec. \ref{sec:ALCHO} we show how to derive analytically the entanglement entropy of a system of coupled harmonic oscillators. We then show the generalisation of such procedure to nonminimally coupled scalar field theories in Sec. \ref{sec:stt}. In Sec. \ref{sec:BHM} we present the quantum black hole metrics and we compute the entanglement entropy in such spacetimes. Conclusion an future perspectives are discussed in Sec. \ref{sec:concl}. In Appendix \ref{sec:RstarEstimation} we present four arguments to estimate the radius at which quantum effects should become dominant. Appendix \ref{sec:num_consid} contains some numerical considerations, while Appendix \ref{sec:cohe_quant} and \ref{sec:bound} review the derivations respectively of the coherent and the corrected-Hayward metrics.  Planck units $c=G=\hbar=k_B=1$ are used throughout the paper.

\section{Theoretical set up}\label{sec:ALCHO}

The Lagrangian of a system of classical coupled harmonic oscillators is given by \cite{qft2,Katsinis:2017qzh}
\begin{equation}\label{eq:HCLMatrix}
    L = \frac{1}{2}M\dot{\boldsymbol{\eta}}^T\boldsymbol{I}\dot{\boldsymbol{\eta}}
        - \frac{1}{2} \boldsymbol{\eta}^T\boldsymbol{K}\boldsymbol{\eta},
\end{equation}
where $\boldsymbol{\eta}\in \mathbb{R}^N$ is the vector containing the displacement from the equilibrium position of every oscillator in the system, $M$ is the mass of one oscillator and $\boldsymbol{K}$ is the matrix that describes the coupling structure of the system. Let $\boldsymbol{y} = \boldsymbol{U}\boldsymbol{\eta}$, where $\boldsymbol{U}$ is the orthonormal matrix that brings $\boldsymbol{K}$ into the diagonal matrix $\boldsymbol{\Gamma}$. With this transformation, the diagonalized Hamiltonian reads
\begin{equation}\label{eq:HCHamiltonian}
    \begin{aligned}
        H&=\frac{1}{2}M\dot{\boldsymbol{y}}^T\boldsymbol{I}\dot{\boldsymbol{y}}+\frac{1}{2} \boldsymbol{y}^T\boldsymbol{\Gamma}\boldsymbol{y}=\\
        &=\sum_{\alpha=0}^{N}\frac{1}{2}M\dot{y}_{\alpha}^2+\frac{1}{2}\Gamma_{\alpha} y_{\alpha}^2.
    \end{aligned}
\end{equation}

We now interpret this Hamiltonian quantum mechanically with the appropriate substitutions.
The eigenvalue equation for the Hamiltonian, in position space, is then given by
\begin{equation}
    \left[\sum_{\alpha=0}^{N}\frac{\hat{p}_\alpha^2}{2M}+\frac{1}{2}M\omega_\alpha^2\hat{y}_{\alpha}^2\right]\Psi=E\Psi,
\end{equation}
where $\omega_\alpha^2=\Gamma_{\alpha}/M$. Since the Hamiltonian in this form represents a system of non interacting harmonic oscillators, it is separable as $\sum_{\alpha} \hat{h}_\alpha$.
It is then straightforward to compute the ground state of the system, which is given by
\begin{equation}
    \Psi=\left(\frac{M}{\pi}\right)^{\frac{N}{4}}\mathrm{Det}(\boldsymbol{\omega})^{\frac{1}{4}}e^{-\frac{M}{2}\boldsymbol{y}^T\boldsymbol{\omega}\boldsymbol{y}},
\end{equation}
where we have introduced the diagonal matrix $\boldsymbol{\omega}$, which has $\omega_\alpha$ as eigenvalues. The ground state density matrix is therefore
\begin{equation}
    \rho(\boldsymbol{y},\boldsymbol{y'})=\left(\frac{M}{\pi }\right)^{\frac{N}{2}}\mathrm{Det}(\boldsymbol{\omega})^{\frac{1}{2}}e^{-\frac{M}{2}\left(\boldsymbol{y}^T\boldsymbol{\omega}\boldsymbol{y}+\boldsymbol{y}'^T\boldsymbol{\omega}\boldsymbol{y}'\right)}.
\end{equation}
We assume now to trace out $n$ oscillators belonging to a definite spatial region, which we will call the inside region. We then write
\begin{equation} \label{outstate}
    \rho_O=\left(\frac{M}{\pi }\right)^{\frac{N}{2}}\mathrm{Det}(\boldsymbol{\Omega})^{\frac{1}{2}}\int \prod_{i=1}^{n}\mathrm{d}\eta_i e^{-\frac{M}{2}\left(\boldsymbol{\eta}^T\boldsymbol{\Omega}\boldsymbol{\eta}+\boldsymbol{\eta}'^T\boldsymbol{\Omega}\boldsymbol{\eta}'\right)},
\end{equation}
where $\rho_O$ denotes the density matrix of the outside system. Furthermore, in Eq. (\ref{outstate}) we have defined
\begin{equation} \label{coupmat}
\boldsymbol{\Omega}=\boldsymbol{U}\boldsymbol{\omega}\boldsymbol{U}^T=\sqrt{\boldsymbol{K}/M}
\end{equation}
and introduced vectors $\boldsymbol{\eta}=(\eta_1,\hdots,\eta_n,\eta_{n+1},\hdots,\eta_{N})$ and $\boldsymbol{\eta}'=(\eta_1,\hdots,\eta_n,\eta_{n+1}',\hdots,\eta_{N}')$.

The integral Eq. (\ref{outstate}) can be computed by introducing the following definitions:
\begin{equation}
    \begin{aligned}
    &\hspace{2.7cm}\boldsymbol{\Omega}=
    \begin{pmatrix}
        \boldsymbol{A}&\boldsymbol{B}\\
        \boldsymbol{B}^T&\boldsymbol{D}
    \end{pmatrix},\\
    &\boldsymbol{\gamma}=-\boldsymbol{B}^T\boldsymbol{A}^{-1}\boldsymbol{B}/2+\boldsymbol{D},\qquad \boldsymbol{\beta}=\boldsymbol{B}^T\boldsymbol{A}^{-1}\boldsymbol{B}/2.
    \end{aligned}
\end{equation}
We can then compute the entanglement entropy relative to the state $\rho_O$, which turns out to be
\begin{equation}\label{eq:ent_entropy_formula}
    S(\rho_O)=\sum_{i=0}^{N-n}\left[-\ln(1-\xi_i)-\frac{\xi_i}{1-\xi}\ln(\xi_i)\right],
\end{equation}
where
\begin{equation}
   \xi_i=\frac{\beta'_i}{1+\sqrt{1-{\beta'}_i^2}},
\end{equation}
and ${\beta'}_i$ are eigenvalues of $\boldsymbol{\gamma}^{-1}\boldsymbol{\beta}$.

\subsection{Area law for nonminimally coupled scalar field}\label{sec:stt}

We now consider a scalar field propagating in a background spacetime described by the metric tensor $g_{\mu\nu}$. The corresponding action reads
\begin{equation}
    S=\int\d^4x\sqrt{-g}\left(\partial_{\mu}\phi\partial^{\mu}\phi-\left(m^2+\xi R\right)\phi^2\right),
\end{equation}
where $R$ is the Ricci scalar curvature, $\xi$ the field-curvature coupling constant, $m$ is the field mass and $g$ the determinant of the metric. The coupling considered is a Yukawa-like interaction, which has found interesting applications both in cosmological and gravitational contexts (see e.g. \cite{faki, mae, blm1}). We specify to a spherically symmetric Schwarzschild-like metric, whose line element reads
\begin{equation} \label{spheline}
    ds^2=f(r)\d t^2-\frac{\d r^2}{f(r)}-r^2(\d\theta^2+\sin^2{\theta}\d\varphi^2),
\end{equation}
where $f$ is a generic function of $r$, which can have one or more zeroes. In order to discretise the scalar field degrees of freedom, we consider the usual expansion in real spherical harmonics
\begin{equation} \label{sphexp}
    \phi(\boldsymbol{x})=\sum_{lm}^{\infty}\phi_{lm}(t,r)Z_{lm}(\theta,\varphi).
\end{equation}
Starting from Eq. (\ref{sphexp}), we can trace back the scalar field to a system of coupled harmonic oscillators, whose Hamiltonian can be expressed in the form of Eq. (\ref{eq:HCHamiltonian}). A suitable strategy consists in employing  Lema\^itre coordinates \cite{Das:2007vj}
\begin{equation}
    \begin{aligned}
        \zeta&=t+\int\d r\frac{1}{f(r)\sqrt{1-f(r)}},\\
        \chi&=t\pm \int\d r \frac{\sqrt{1-f(r)}}{f(r)}.
    \end{aligned}
\end{equation}
In particular, by fixing the Lema\^itre time $\chi=0$ we can write the scalar field Hamiltonian as \cite{Das:2007mj}
\be \label{hamsum}
H= \sum_{lm} H_{lm}(0),
\ee
with
\begin{align} \label{hamexp}
H_{lm}(0)= \frac{1}{2} &\int_0^\infty dr \bigg[ \frac{\pi_{lm} r^{-2}}{1-f(r)} + r^2 \left( \partial_r \phi_{lm}\right)^2 \notag \\
&+ l(l+1)\phi_{lm}^2+ r^2\left(m^2+\xi R\right) \phi_{lm}^2\bigg].
\end{align}
The canonical momenta $\pi_{lm}$ satisfy the Poisson brackets
\be \label{pois}
\{ \phi_{lm}(r), \pi_{lm}(r^\prime) \}= \sqrt{1-f(r)} \delta(r-r^\prime)
\ee
and, by means of the canonical transformation
\be \label{cantrasf}
\pi_{lm} \rightarrow r \sqrt{1-f(r)} \pi_{lm},\ \ \ \ \phi_{lm} \rightarrow \frac{\phi_{lm}}{r},
\ee
we arrive at
\begin{equation} \label{radham}
H_{lm}(0)= \frac{1}{2} \int_0^\infty dr \bigg \{\pi_{lm}^2(r)+ r^2 \bigg[ \frac{\partial}{\partial r}\left( \frac{\phi_{lm}}{r} \right) \bigg]^2 + \mathcal F_l \phi_{lm}^2 \bigg \},
\end{equation}
where we have introduced the compact notation
\begin{equation}
\mathcal F_l\equiv \left( \frac{l(l+1)}{r^2}+m^2+ \xi R \right).
\end{equation}
The above equation shows that field-curvature coupling adds a position-dependent term to the field Hamiltonian. In particular, for a metric tensor described by Eq. (\ref{spheline}), the scalar curvature is derived from the lapse function $f(r)$ as
\be \label{ricscal}
R(r)=\frac{2-2f(r)-4rf^\prime(r)-r^2 f^{\prime \prime}(r)}{r^2}.
\ee
Once decomposed the field in real spherical harmonics, we still have to deal with the continuous radial coordinate $r$. In the following, we show how the theory is regularized by introducing a lattice of spherical shells, exploiting the symmetries of the system.

\subsection{Discretising the Hamiltonian}

The only continuous variable left in Eq. (\ref{radham}) is the radial coordinate $r$. To further proceed, we introduce an ultraviolet cutoff length, usually denoted by $a$, and discretise $r$ as $r_j=aj$, where $j$ is a positive integer \cite{qft2}. The introduced cutoff length, $a$, represents an additional free parameter in the model. For our analysis, it seems natural to set $a=1$, since we aim to focus on quantum effects at Planck lengthscales. Additionally, we impose the maximum value of $j$ to be $N$, thus fixing also an infrared cutoff associated with the total size of the system. The fully discretised Hamiltonian then reads\footnote{Since we employ forward discrete derivatives, the total size of the system is $A=(N+1)a$ in our case, where $A$ acts as the infrared cutoff.}
\begin{align} \label{discham}
H= &\frac{1}{2a} \sum_{lm} \sum_{j=1}^N \bigg[ \pi_{lm,j}^2 + \left( j+\frac{1}{2} \right)^2  \left( \frac{\phi_{lm,j+1}}{j+1}-\frac{\phi_{lm,j}}{j} \right)^2  \notag \\
&+ \left( \frac{l(l+1)}{j^2}+ (m^2+\xi R_j)a^2  \right) \phi_{lm,j}^2 \bigg],
\end{align}
where $\phi_{lm,j} \equiv \phi_{lm}(r_j)$ and $\pi_{lm,j} \equiv \pi_{lm} (r_j)$ and $R_j=R(r_j)$. Since in Eq. (\ref{discham}) there is no dependence on the $m$ index, we can rewrite the Hamiltonian of the system as
\be \label{haml}
H= \sum_{l=0}^\infty (2l+1) H_l,
\ee
where $H_l$ is now in the form of Eq. (\ref{eq:HCLMatrix}). Therefore, the scalar field behaves as a system of coupled harmonic oscillators, where the coupling matrix $\boldsymbol{K}$ defined in Eq. (\ref{eq:HCLMatrix}), is now given by
\begin{align} \label{kmat}
\boldsymbol{K}_{ij}=  & \left( \frac{l(l+1)}{i^2}+(m^2+\xi R_i) a^2  \right) \delta_{ij}  \notag \\
&+\frac{1}{i^2} \bigg[ \frac{9}{4} \delta_{i1} \delta_{j1} +\left(N-\frac{1}{2} \right)^2 \delta_{iN} \delta_{jN} \notag \\
&\ \ \ \ \ \ \ \ \ +\left( \left(i+\frac{1}{2} \right)^2+ \left(i-\frac{1}{2}  \right)^2   \right) \delta_{ij\ (i \neq 1,N)} \bigg] \notag \\
& -\left[\frac{(j+1/2)^2}{j(j+1)} \right] \delta_{i,j+1}  -\left[ \frac{(i+1/2)^2}{i(i+1)} \right] \delta_{i+1,j}. \notag \\
\end{align}
The last two terms represent nearest-neighbour interactions, which originate from the spatial derivatives of the field and show the local character of the Hamiltonian.

In the next section, we focus on quantum black hole spacetimes and we proceed in concretely computing the entanglement entropy, which is given by
\begin{equation}\label{eq:ent_entropy_nl}
    \mathcal{S}(n,N)= \sum_l (2l+1) \mathcal{S}_l(n,N),
\end{equation}
where $\mathcal{S}_l(n,N)$ is computed through Eq. (\ref{eq:ent_entropy_formula}), using Eq. (\ref{kmat}) as the coupling matrix.

\section{Area law in quantum black holes}\label{sec:BHM}

Here we apply the formalism presented in the previous section to relevant quantum black hole proposals, with the aim of identifying possible quantum effects in entanglement generation for discrete scalar field theories. We also provide a direct comparison with Hayward-like regular black holes, which correctly reproduce Schwarzschild geometry at sufficiently large distances.

\subsection{Quantum-corrected Schwarzschild metric} \label{sec:quasann}

The quantum-corrected Schwarzschild lapse function proposed in Ref. \cite{Binetti:2022xdi} reads
\begin{equation}\label{eq:QuantumLapseF}
    f(r)=1-\frac{r_s}{r}\sum_{n=0}^{\infty}\frac{\Omega_n}{[d(r)]^{2n}},
\end{equation}
where  $r_s=2M$ is the Schwarzschild radius and $M$ the black hole mass, $r$ is the radial coordinate and the parameters $\Omega_n$, with $\Omega_0=1$, describe the underlying theory of gravity, which is not fixed a priori. Furthermore, the function $d(r)$ is the proper distance from the center of the black hole. The definition of Eq. (\ref{eq:QuantumLapseF}) is therefore circular and requires an iterative computation. To address this, we consider the first two terms in the sum and employ the Schwarzschild proper distance $d_0(r)$ instead of the full proper distance\footnote{The new proper distance computed with this metric will be denoted as $d_1(r)$. } $d(r)$. Since we are going to compute the entanglement entropy by tracing over a volume close to the origin, we are allowed to use the following approximated expression for the proper distance \cite{Binetti:2022xdi}:
\begin{equation}
    d_0(r)=\frac{2}{3}\frac{r^{\frac{3}{2}}}{\sqrt{r_s}}+\mathcal{O}(r^{\frac{5}{2}}).
\end{equation}
The leading order quantum-corrected lapse function near the origin then reads
\begin{equation}\label{eq:LOQLF}
    f(r)=1-\frac{r_s}{r}\left(1+\frac{\Omega_1}{[d_0(r)]^2}\right)
\end{equation}
We remark that Eq. (\ref{eq:QuantumLapseF}) is defined only at distances greater then the Planck length, and it is therefore regular. In particular, the quantum-corrected lapse function can be written as the sum of the classical Schwarzschild contribution and a quantum correction\footnote{A quantum correction due to the trace anomaly effect can be implemented by setting $\Omega_1=-4/9 c_A $, where $c_A$ is a function of the number of particles of each species \cite{Abedi:2015yga}.},
\begin{equation}\label{eq:qua_corrected_lapse}
f(r)=1-\frac{r_s}{r}-\frac{9\Omega_1r_s^2}{4r^4}=f_{S}(r)+f_{Q}(r).
\end{equation}
We notice that, far enough from the origin, the lapse function reduces to the Schwarzschild one, while close to the origin quantum effects dominate. As discussed in Appendix \ref{sec:RstarEstimation}, we can introduce a radius $a_g$ to separate the regime where quantum effects are dominant ($r\ll a_g$) from the region where classical general relativity dominates ($r\gg a_g$). This length is computed through quantum mechanical considerations applied to a Schwarzschild black hole. We can derive a similar quantity, denoted as $r^*$, exclusively using the quantum metric. In the specific case of Eq. (\ref{eq:qua_corrected_lapse}), we can estimate the $r^*$ by the following argument. The quantum-corrected metric consist of three parts: a constant term, a term analogous to the Newtonian potential and a quantum correction term. We seek the radius at which the quantum correction is of the same order of the Newtonian term. This happens at $r^*=(9\Omega_1 r_s/4)^{1/3}$ and thus quantum effects dominate for $r<r^*$, as the correction term surpasses the classical one. Therefore, to find an appropriate value for the quantum theory parameter $\Omega_1$, we solve $r^*=a_g$ obtaining
\begin{equation}\label{eq:quantum_r_star}
    \Omega_1\approx \frac{4}{9}\frac{a_g^3}{r_s},
\end{equation}
The value of $\Omega_1$ thus depends on the size of the black hole and on the chosen form of $a_g(r_s)$, which is further discussed in Appendix \ref{sec:RstarEstimation}.

From Eq. (\ref{eq:LOQLF}), we can obtain an expression for the scalar curvature of Eq. (\ref{ricscal}), which depends only on the quantum corrections to the Schwarzschild metric, namely
\begin{equation}\label{eq:quantum_curvature}
    R(r)=\frac{-2f_Q(r)-4rf_Q'(r)-r^2f_Q''(r)}{r^2}=\frac{27\Omega_1 r_s^2}{2r^6}
\end{equation}
For $\Omega_1>0$, such quantum-corrected Schwarzschild spacetime has a single horizon located at
\begin{equation}
    r_+=r_s \left[1+\alpha+\mathcal{O}(\alpha^{\frac{3}{2}})\right],\quad\mathrm{with}\quad \alpha=\frac{4|\Omega_1|}{\pi r_s^2}
\end{equation}
By increasing the black hole mass, or by diminishing the parameter describing quantum effects, such horizon can be made arbitrarily close to the classical one.

For $\Omega_1<0$, the horizon shrinks with respect to the classical one
\begin{equation}
    r_+=r_s \left[1-\alpha+\mathcal{O}(\alpha^{\frac{3}{2}})\right],
\end{equation}
and the considerations made for positive $\Omega_1$ still hold. A notable difference in this case is that the black hole can develop one or more internal horizons, which are highly dependent on the underlying quantum theory
\begin{equation}
    r_-=\frac{r_s}{2}\left(\frac{9\pi}{2}\right)^\frac{1}{3}\left[\alpha^\frac{1}{3}+\frac{1}{5}\left(\frac{\pi^2}{6}\right)^{\frac{1}{3}}\alpha^\frac{2}{3}+\mathcal{O}(\alpha)\right].
\end{equation}
The presence or absence of the internal horizons depends strongly on the size and sign of the higher order quantum corrections, e.g $\Omega_2$, \cite{Binetti:2022xdi}.

\begin{figure}[H]
    \centering
    \includegraphics[width=1\columnwidth,clip]{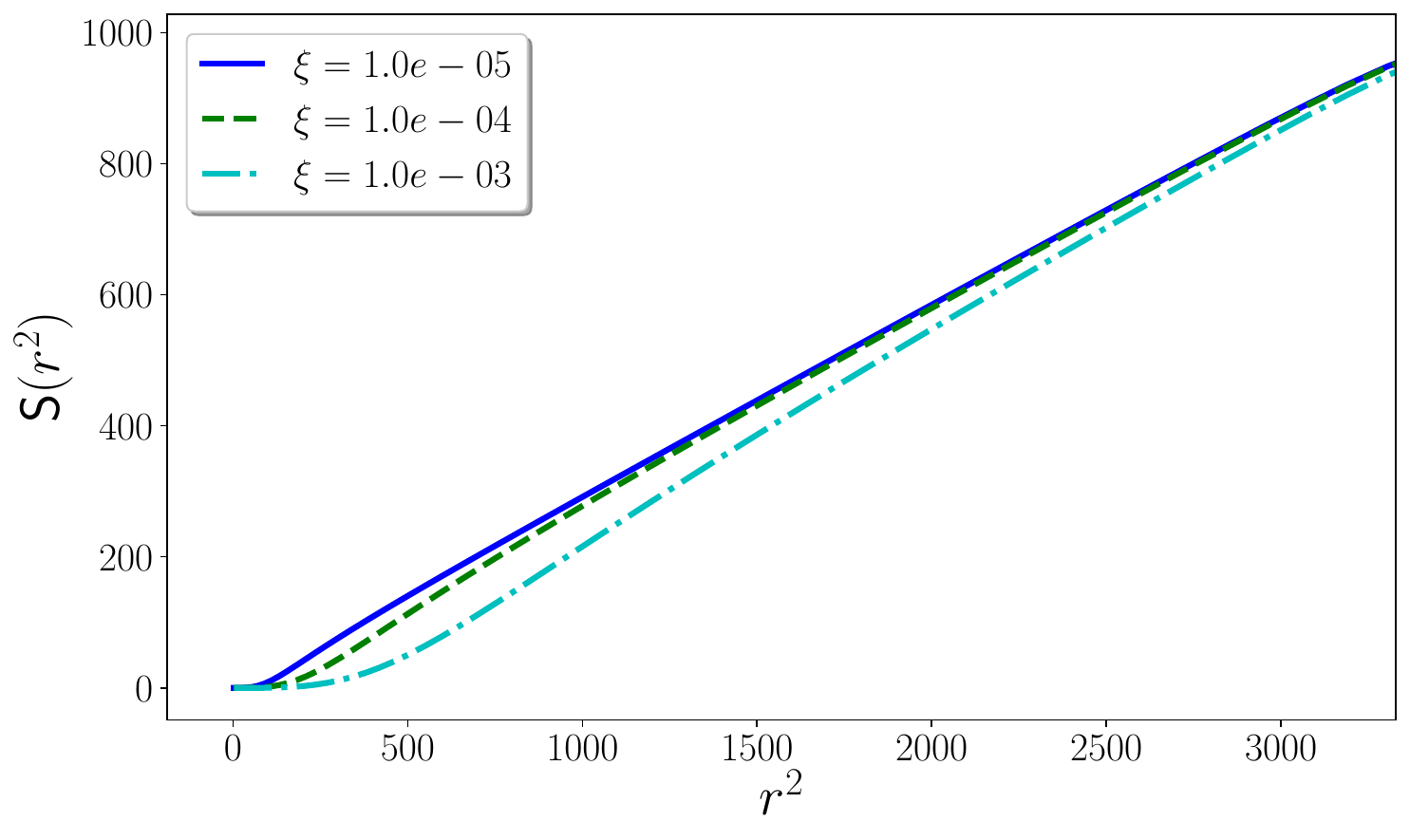}
    \caption{Entanglement entropy in the quantum black hole spacetime described by Eq. (\ref{eq:qua_corrected_lapse}), for different values of the coupling constant $\xi$. The parameters used are $a=1$, $m=8.2\cdot 10^{-20}$, $n_{\mathrm{min}}=1$, $N=60$, $\Omega_1=0.22$, $r_s=2\cdot 10^{5}$, $l_{\mathrm{max}}=10^3$. The calculation of $a_g$ is performed using Eq. (\ref{eq:trace}).}
    \label{fig:eeQ}
\end{figure}
Since we compute the entanglement entropy using Lema\^itre coordinates, we shall restrict ourselves to positive values of $\Omega_1$, in order to avoid the additional complexity of having internal horizons and to remain consistent with Eq. (\ref{eq:quantum_r_star}). The numerical results are presented in Fig. \ref{fig:eeQ}, which illustrates the impact of the quantum corrections on the scaling of entanglement entropy. Some details about the numerical simulations can be found in Appendix \ref{sec:num_consid}. As mentioned above, the metric is defined for $r\geq1$ and we therefore start to trace out the oscillators form the first shell, $n_{\mathrm{min}}=1$. The last shell is instead given by $n_{\mathrm{max}}\equiv n_{\mathrm{min}}+N$ ( see Appendix \ref{sec:num_consid}).

From Eq. (\ref{eq:quantum_curvature}), we also find that $R \propto r^{-6}$, and this results in a significant decrease of the entropy for small radii, since the curvature acts to modify the mass of the field in a coordinate-dependent manner. Inspired by the form of Eq. (\ref{discham}), we define the effective field mass as
\begin{equation}\label{eq:eff_mass}
    m^2_{\mathrm{eff}}(r)\equiv m^2+\xi R(r)= m^2 + \frac{27}{2}\frac{\xi\Omega_1 r_s^2}{r^6}.
\end{equation}
We notice that the effective mass functionally depends on $\xi\Omega_1$ and not on each parameter individually, so varying either parameter produce identical results. As illustrated in Appendix \ref{sec:num_consid}, Fig. \ref{fig:flat_mass}, increasing the field mass leads to an overall decrease in the slope of the entanglement entropy. It is evident that in the high field mass limit, $m \rightarrow \infty$, the entropy tends to zero, since it is a decreasing function of the scalar field mass. By examining Eq. (\ref{eq:eff_mass}), we observe that, at $r=1$, the effective mass reaches its maximum, resulting in a significant suppression of the entanglement entropy. Interestingly, when we move sufficiently far from the origin\footnote{With the parameters values reported in Fig. (\ref{fig:eeQ}), the condition is satisfied for $r^2 \gg 400$.}, such that $m^2_{\mathrm{eff}}(r) \ll 1$, the entanglement entropy recovers the linear behaviour of minimally coupled scenarios\footnote{A brief motivation for the choice $m^2_{\mathrm{eff}}(r) \ll 1$ is given in Appendix \ref{sec:num_consid_mass}.}.
Therefore, at large distances from the origin, the area law is approximately restored for such quantum black hole, even in the presence of a Yukawa-like nonminimal coupling.

\subsection{Coherent quantum black holes} \label{sec:quantcoh}

The quantum-corrected metric analysed in the previous section does not rely on any particular assumption on the underlying quantum theory. As an alternative proposal, in Ref. \cite{casadio2022geometry} it is assumed that the black hole's gravitational potential arises from the mean value of a massless scalar field on a particular coherent state. A brief overview of the derivation can be found in Appendix \ref{sec:cohe_quant}. The resulting lapse function is given by
\begin{equation}\label{eq:cohLapse}
    f(r)=1-\frac{r_s}{r}\frac{2}{\pi}\mathrm{Si}\left(\frac{r}{R_S}\right),
\end{equation}
where $R_S$ is the radius within which the \virgolette{quantum} mass is localised, $r_s$ is the Schwarzschild radius, and $\mathrm{Si}(x)$ is the sine integral function defined by $\mathrm{Si}(x)=\int_0^x \frac{\sin t}{t}\mathrm{d}t$.

The scalar curvature in a spacetime described by Eq. (\ref{eq:cohLapse}) is found to be
\begin{equation}\label{eq:cohCurv}
    R(r)=\frac{2r_s}{\pi R_S}\frac{r\cos (\frac{r}{R_S})+R_S \sin (\frac{r}{R_S})}{ r^3 }.
\end{equation}
Since we are interested in the region close to the origin, we can assume $r\ll R_S$, so to obtain the simplified expression
\begin{equation}\label{eq:coh_approx_curv}
    R(r)=\frac{4r_s}{\pi R_S}\frac{1}{ r^2 }.
\end{equation}
We underline that the point $r=0$ represents an integrable singularity within such quantum-corrected approach. Thus, despite the divergence of some geometric invariants, matter still experiences finite forces \cite{casadio2022geometry}.

In order to find a physically motivated value for $R_S$, we employ the same strategy used before. For this metric, the radius that separate the quantum behaviour from the classical one is $r^*\approx R_S$, which results in $R_S\approx a_g$.

In this case, it is difficult to show in a single plot the deviations due to the curvature coupling together with the linear behaviour. This is because to do that, the two following conditions should hold
\begin{equation}\label{eq:alfa_constrains}
   \alpha\gg 1 \quad \mathrm{and}\quad \frac{\alpha}{ N^2 }\ll 1 \quad \mathrm{where} \quad \alpha= \frac{2\xi r_s}{\pi R_S}
\end{equation}
These inequalities imply $m^2_{\mathrm{eff}}(r=1)\gg1$ and $m^2_{\mathrm{eff}}(r=N)\approx 1$.
The first inequality sets the value of $\alpha$. The only way to satisfy the second inequality is by considering a large $N$. However, due to the complexity of the numerical computation, we cannot exceed $N\approx 100$ on our machine. Nevertheless, we know that the area law behaviour must reestablish for large $r$, since the curvature in Eq. (\ref{eq:cohCurv}) tends to zero as $r \rightarrow \infty$.

\begin{figure}[H]
    \centering
    \includegraphics[width=1\columnwidth,clip]{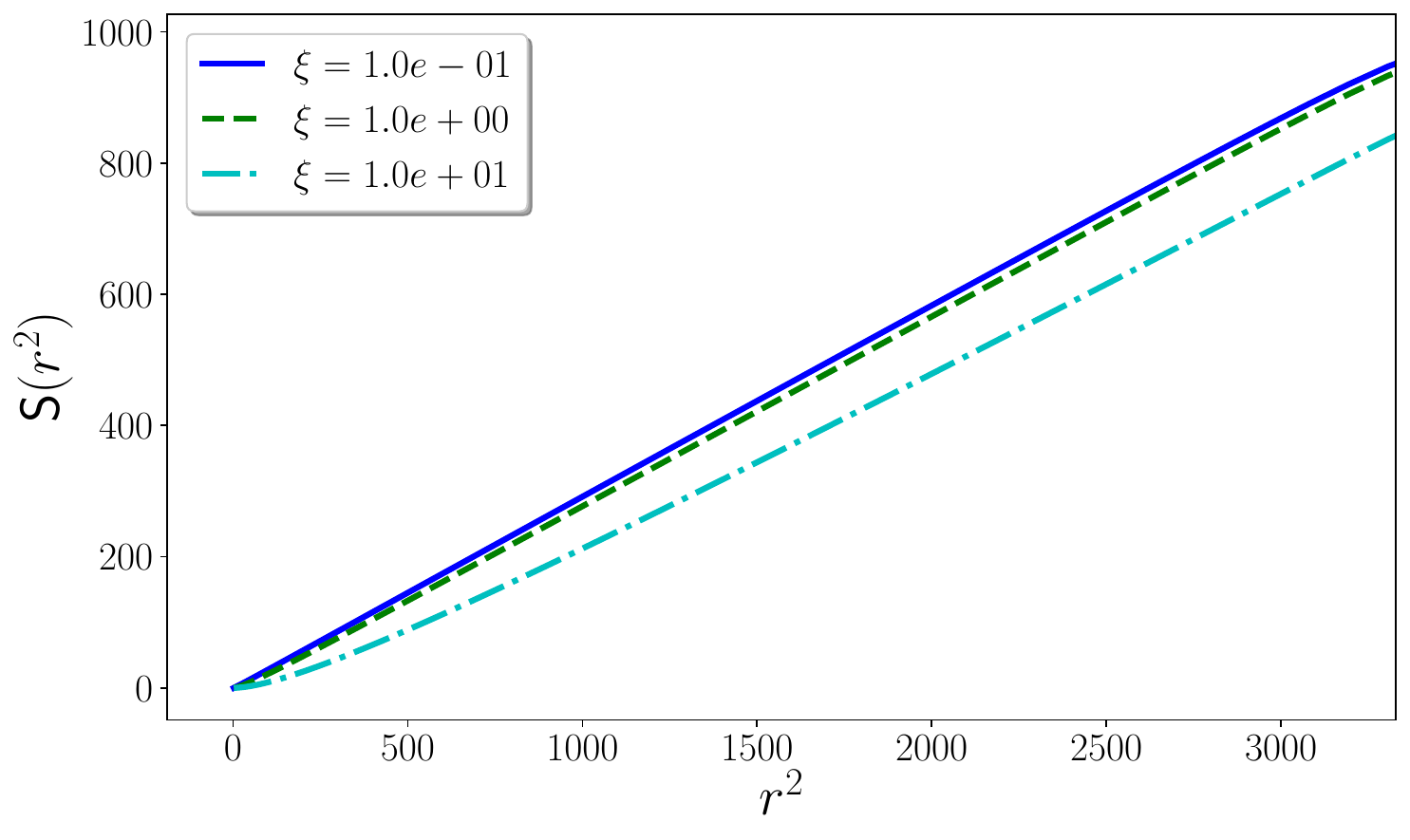}
    \caption{Entanglement entropy in the coherent quantum black hole spacetime described by Eq. (\ref{eq:cohLapse}), for different values of the coupling constant $\xi$. The parameters used are $a=1$, $m=8.2\cdot10^{-20}$, $n_{\mathrm{min}}=1$, $N=60$, $R_S=5\cdot10^{37}$, $r_s=2\cdot 10^{39}$ $l_{\mathrm{max}}=10^3$. To compute $a_g$, we used Eq. (\ref{eq:corpuscular}). }
    \label{fig:eeCQ}
\end{figure}
The numerical results are shown in Fig. \ref{fig:eeCQ}. The quantum oscillations are not visible since the curvature essentially reduces to Eq. (\ref{eq:coh_approx_curv}) close to the origin.  We have chosen to set $\alpha \gg 1$ to make the deviation from the area law clearer. We observe more pronounced deviations as $r \rightarrow 0$, while the difference with respect to the free-field area law seems to become constant as the radial distance increases. This behaviour is qualitatively similar to the one observed for the quantum-corrected Schwarzschild metric of Eq. (\ref{eq:QuantumLapseF}), with the difference that we cannot observe here the \virgolette{catch up} due to the constraints of Eq. (\ref{eq:alfa_constrains}).

\subsection{Comparison with regular black holes}

\subsubsection{Hayward}
The Hayward metric offers a simple description of a regular black hole. It serves as a minimal model that is both regular and static, exhibiting spherical symmetry and asymptotic flatness. This metric provides a framework for testing the behaviour of entanglement entropy in a scenario involving a regular black hole, as discussed in \cite{qbh6,Belfiglio:2023sru}. The corresponding lapse function is defined as
\begin{equation} \label{hayw}
f(r)=1-\frac{r_s r^2}{r^3+r_s b^2},
\end{equation}
where $r_s$ is approximately the outer\footnote{There is also an inner horizon at $r_-\approx b$.} horizon radius and $r$ is the radial coordinate. Near the origin, the lapse function reproduces a de Sitter core, with $b^2=1/\Lambda^2$, where $\Lambda$ represents the cosmological constant in this region. Given that $b$ is intended to define the scale at which quantum-gravitational effects become significant, we will set its value to $a_g$.

The scalar curvature of the Hayward metric is given by the expression
\begin{equation}\label{haycur}
R(r)= \frac{6 b^2 r_s^2 \left( 2 b^2 r_s-r^3 \right)}{\left( b^2 r_s+r^3 \right)^3}.
\end{equation}
This expression is coordinate-dependent, implying deviations from the area law. As $r\rightarrow 0$, the effective mass, Eq. (\ref{eq:eff_mass}), can be expressed as
\begin{equation}\label{eq:hay_m_eff}
m_{\mathrm{eff}}^2(r)\approx m^2 +\xi\frac{12}{b^2}. 
\end{equation}
Since in this limit we are well inside the quantum core of the metric, we are in a de Sitter space characterised by a constant curvature. The effect is to enlarge the field mass by a constant value. This suggest an area law behaviour, with a lower angular coefficient compared to the free-field case\footnote{Further details are discussed in Appendix \ref{sec:num_consid_mass}.}.

\begin{figure}[H]
    \centering
    \includegraphics[width=1\columnwidth,clip]{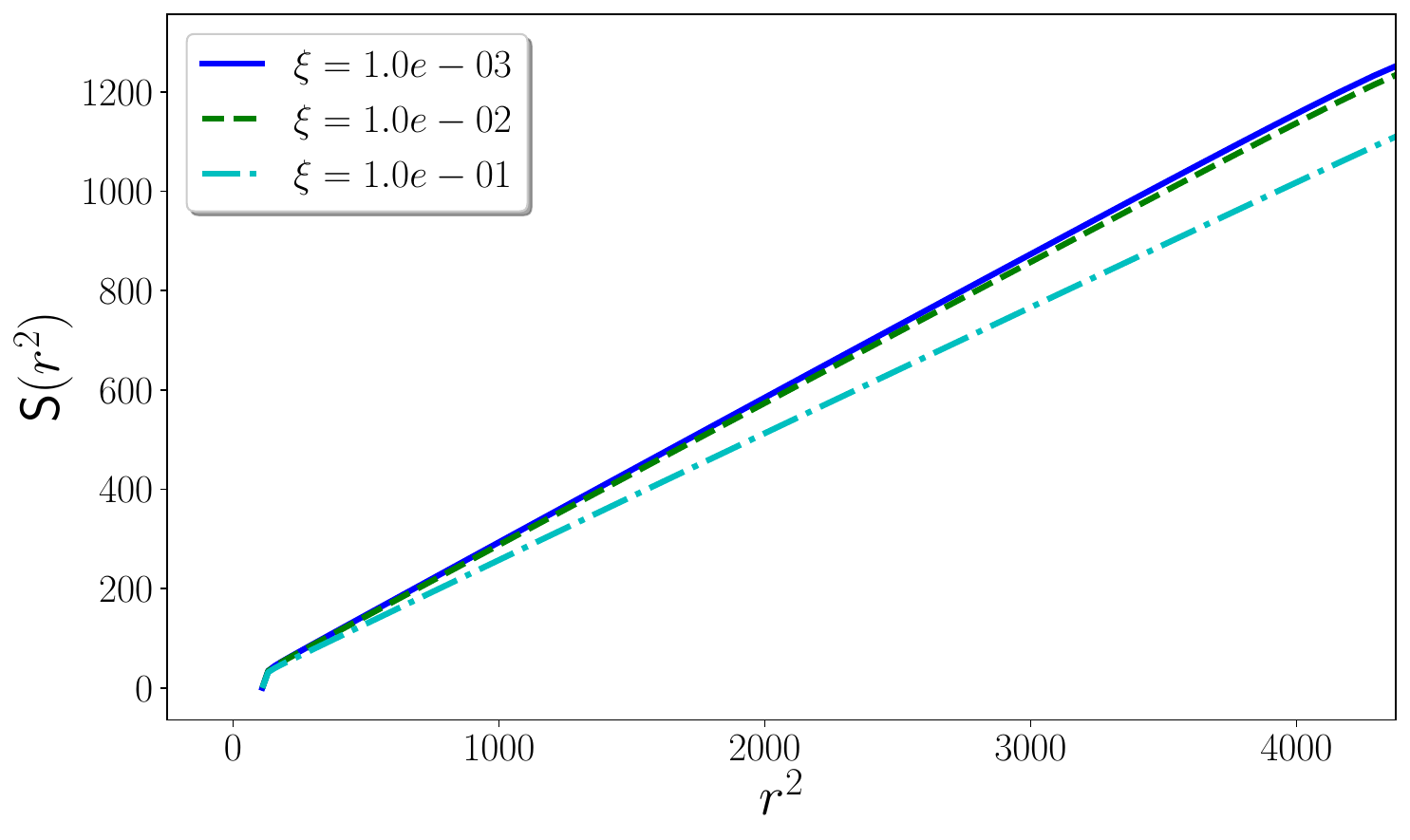}
    \caption{Entanglement entropy in Hayward spacetime, Eq. (\ref{hayw}), for different values of the coupling constant to curvature $\xi$. The parameters used are $a=1$, $b=3$, $n_\mathrm{min}=10$, $N=60$, $l_\mathrm{max}=10^3$, $m=8.19\cdot10^{-20}$, $r_s=2\cdot 10^{44}$. We compute $a_g$ according to Eq. (\ref{eq:heisenberg}). }
    \label{fig:hayw}
\end{figure}
Results are shown in Fig. \ref{fig:hayw}. We compute the area law starting from the $10$-th shell, in order avoid crossing the internal horizon, where the approach described in Sec. \ref{sec:ALCHO} would break down. Thus, at $r^2=100$ the entanglement entropy must be zero and this results in the observed edge effect.

Thus the area law holds in Hayward black hole spacetimes, even if the angular coefficient is modified by the de Sitter constant curvature. This is quite different from what we observed for the previously described quantum metrics, for which the area law is broken near the origin due to the scalar curvature.

\subsubsection{Corrected-Hayward}

More recently, some modifications to regular black hole models have been proposed, asking for example whether such spacetimes can be obtained from an action principle in quantum gravity.  In Ref. \cite{Knorr:2022kqp}, static spherically symmetric metrics are considered, and a lower bound for the power of the asymptotic correction term is obtained\footnote{A simplified review of their argument can be found in Appendix \ref{sec:bound}.}. In particular, it is shown that Hayward solution is not compatible with a stationary local-action principle.  A suitable modification of Eq. (\ref{hayw}) is instead
\begin{equation}\label{eq:qua_hay}
    f(r)=1-\frac{r_s r^4}{r^5+r_s b^4}.
\end{equation}
The scalar curvature for this metric results
\begin{equation}\label{eq:qua_hay_curvature}
    R(r)=\frac{10  b^4  r^2  r_s^2  (-2  r^5 + 3  b^4  r_s)}{(r^5 + b^4  r_s)^3}.
\end{equation}
To quantify the impact of the field-curvature coupling within this model, we expand the curvature near the origin and define the effective mass, Eq. (\ref{eq:eff_mass}), as
\begin{equation}\label{eq:qua_hay_m_eff}
m_{\mathrm{eff}}^2(r)\approx m^2 +\xi\frac{30 r^2}{b^4}.
\end{equation}
This is quite different from the standard Hayward metric, since in this case the curvature approaches zero close to the origin and increases quadratically with the radial distance.
\begin{figure}[H]
    \centering
    \includegraphics[width=1\columnwidth,clip]{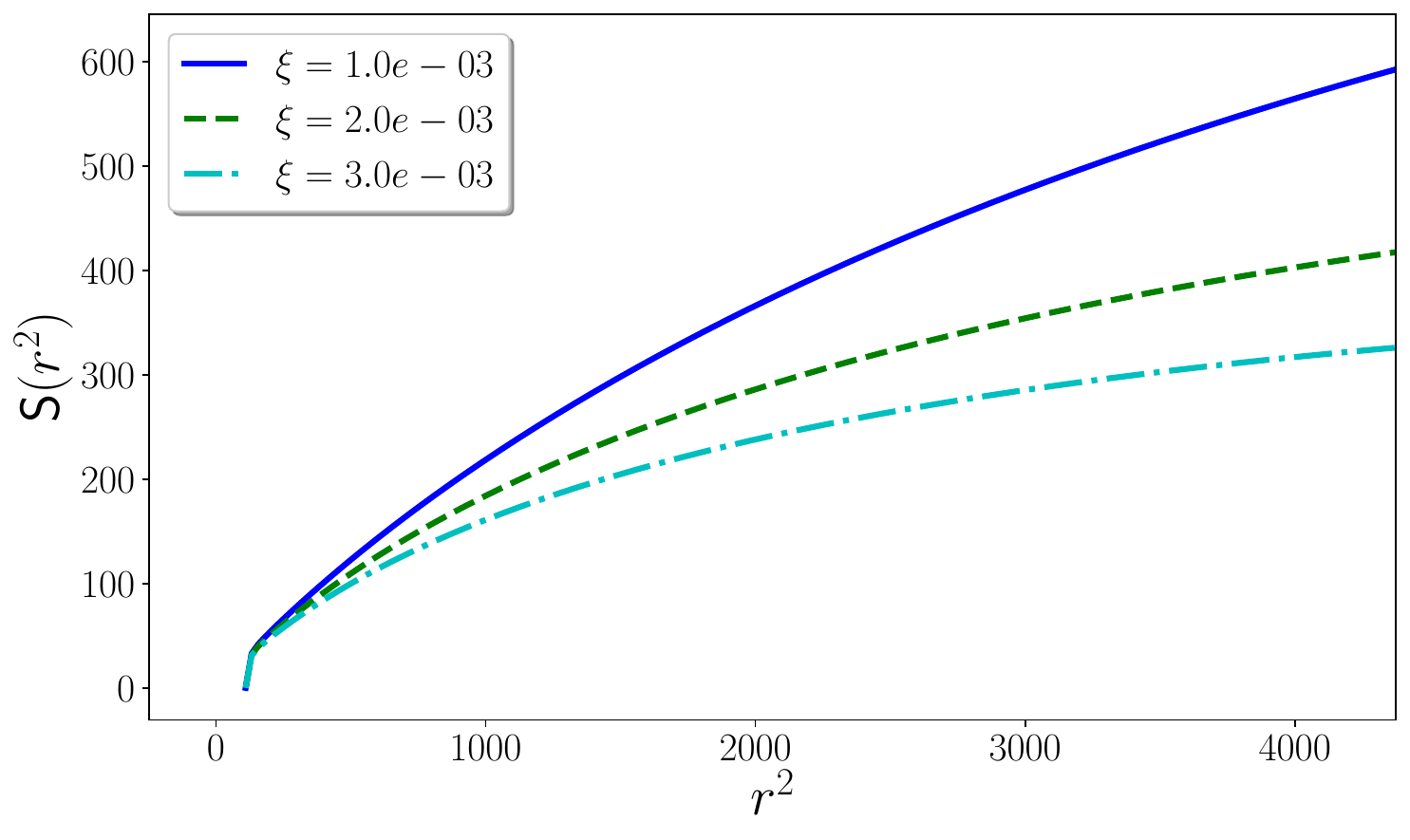}
    \caption{Entanglement entropy in the modified Hayward scenario described by Eq. (\ref{eq:qua_hay}), for different values of the coupling constant $\xi$. The parameters used are $a=1$, $b=3$, $n_\mathrm{min}=10$, $N=60$, $l_\mathrm{max}=10^3$, $m=8.19\cdot10^{-20}$, $r_s=2\cdot 10^{44}$. We compute $a_g$ according to Eq. (\ref{eq:heisenberg}). }
    \label{fig:qua_hayw}
\end{figure}
The corresponding results for the entanglement entropy are shown in Fig. \ref{fig:qua_hayw}. Within this scenario, the free-field area law is restored for sufficiently small distances from the origin. Deviations start to appear when considering larger radii. Clearly, the quadratic dependence of the effective mass on the radius cannot hold indefinitely, since the curvature must vanish asymptotically, in order to recover Schwarzschild geometry.

Surprisingly, this finding is quite the opposite of what was expected based on the quantum metrics, which predicted the largest deviations near the origin.

\section{Discussion on minimum size black holes} \label{Sezdisc}

The quantum black hole metrics presented in Sec. \ref{sec:quasann} and \ref{sec:quantcoh} are characterised by the presence of at least one free parameter, which in principle should quantify deviations from the classical behaviour.
In particular, for what concerns Ref. \cite{casadio2022geometry}, if the radius at which quantum effects dominate approaches the Schwarzschild radius, the classical black hole picture gets disrupted. This happens because the quantum oscillations of the coherent quantum black hole metric in Eq. (\ref{eq:cohLapse}) become large when approaching the horizon. It may be that quantum oscillations completely tear apart the black hole remnant at the end of Hawking evaporation. This is in accordance with our result Eq. (\ref{eq:heisenberg}) and the result from trace anomaly Eq. (\ref{eq:trace}), as when $a_g>r_s$ the quantum effect causes the particles to \virgolette{hover} outside the horizon. This quantum effect could compromise the stability of the black hole.

We present the expressions for $r_s^m$, which represent the minimum Schwarzschild radius, according to the four strategies to compute the radius at which quantum effects dominate, discussed in Appendix \ref{sec:RstarEstimation}:
\begin{align}
    r_s^m &\approx \frac{1}{m} \qquad\qquad\qquad \text{(Heisenberg uncertainty)} \label{eq:heisenberg_rsm} \\
    r_s^m &\approx \left(\frac{c_A}{2}\right)^{3/2} \qquad\quad \,\text{(Trace anomaly)} \label{eq:trace_rsm} \\
    r_s^m &\approx \frac{\chi \xi}{\sqrt{g_*}} \qquad\qquad\quad \text{(Asymptotic safety)}\label{eq:asym_rsm}\\
    r_s^m &\quad \text{is undefined} \qquad \text{(Corpuscular scaling)} \label{eq:corpuscular_rsm}
\end{align}
where $m$ is a fundamental mass scale yet to be determined, $c_A$ is a function of the number of massless fields \cite{Mottola:2010gp}, $\chi,\xi$ are constants of order one and $g^*$ is the fixed point value of the dimensionless gravitational coupling \cite{Pawlowski:2018swz}.
We will now briefly discuss the behaviour of the entanglement entropy in the limiting case $a_g\approx r_s^m$. We first consider the quantum black hole metric described by Eq. (\ref{eq:QuantumLapseF}). Substituting $a_g=r_s^m$ into Eq. (\ref{eq:quantum_r_star}), we obtain $\Omega_1 \approx (r_s^m)^2$. Now, we present the results for each choice of $a_g(r_s)$. Using Eq. (\ref{eq:heisenberg_rsm}), we find $\Omega_1\approx m^{-2}$. Since $m$ is expected to be small, this value is much greater than unity, resulting in a significant suppression of the entanglement entropy. This suggests that quantum effects are more pronounced for a smaller black hole \cite{qbh1}, confirming our intuition. With Eq. (\ref{eq:asym_rsm}), we obtain $\Omega_1 \approx 1$, since from their values of the constants $r_s^m\approx 1$. In this case, the minimum size is around the Planck length and thus it is close to the chosen cutoff length. It is therefore not clear if computing the entanglement entropy through the procedure presented in Sec. \ref{sec:ALCHO} is meaningful. For Eq. (\ref{eq:trace_rsm}), the result depends on the value of $c_A$, which is not straightforward to estimate. Nevertheless, we can expect the radius Eq. (\ref{eq:trace_rsm}) to assume a value intermediate between Eq. (\ref{eq:heisenberg_rsm}) and Eq. (\ref{eq:asym_rsm}). Interestingly, Eq. (\ref{eq:corpuscular_rsm}) does not provide a minimum black hole size.

Finally, we consider the coherent quantum metric Eq. (\ref{eq:cohLapse}). Imposing $r_s=r_s^m=R_S$ yield an effective mass which is independent of $r_s^m$ and thus we would obtain the same entanglement entropy regardless of the choice of $a_g(r_s)$.

\section{Conclusions and perspectives}\label{sec:concl}

In this study, we investigated the influence of nonminimal coupling on the entanglement entropy for a massive scalar field in quantum black hole spacetimes. We focused on static, spherically symmetric solutions and discretised the scalar field following the standard approach, regularizing the theory by means of a lattice of spherical shells.  We then employed Lema\^itre coordinates to transform the discretised scalar field into an effective one-dimensional harmonic system, for which the entanglement entropy in real space can be computed analytically.

We discussed in particular two quantum black hole models. The first one, represented by Eq. (\ref{eq:QuantumLapseF}), aims to parameterise quantum effects through a power series expansion that, in our approximation, depends on a free parameter describing the underlying quantum gravity theory, which is left unspecified. We computed the entanglement in this curved spacetime and observed significant deviations from the free-field area law at small radii. These deviations were understood in terms of the position-dependent effective mass, due to the presence of field-curvature coupling. The second quantum black hole metric, described by Eq. (\ref{eq:cohLapse}), attempts to interpret the gravitational potential as the mean value of a massless scalar field on a coherent state. In this case as well, we observed deviations from the area law at small radii, which were qualitatively similar to the results obtained for the first model.

We also noticed that both quantum metrics depend on a free parameter, associated with the scales at which quantum gravity becomes relevant. To estimate the parameter's value, we explored four different methods and  quantified the radius at which quantum effects become dominant. The first method relies on a simple physical argument based on the assumption that quantum effects are dominated by the Heisenberg indeterminacy principle, preventing matter from collapsing into a point. The other three methods are respectively based on the trace anomaly, asymptotic safety, and graviton corpuscular scaling. Subsequently, we determined the analogous radius using the metrics and ensured their coincidence.

We then computed the entanglement entropy scaling in regular black hole spacetimes. We first considered the Hayward metric, often regarded as the prototype of a regular metric. We found that near the origin, the area law is preserved, but the angular coefficient is not equal to the free-field one. Next, we explored a modification of the Hayward metric that is asymptotically consistent with the least action principle. In such corrected-Hayward metric, very close to the origin, we observed a free-field area law behaviour and deviations at larger radii. Therefore, we observed that entanglement entropy can have very different behaviours in quantum black hole metrics compared to regular metrics.

As future developments, we plan to further investigate the standard strategy for computing the entanglement entropy. This involves overcoming the limitation imposed by the Lema\^itre coordinates, which hinder us from crossing horizons in our computations. Additionally, the impact of this coordinate change on the entanglement entropy is not yet clear. Another goal is to enhance our numerical methods to extend the range within which the entanglement entropy can be computed within a reasonable time-frame. The primary issue with the current discretisation scheme is that, to trace out a finite number of harmonic oscillators, we must perform an infinite sum on the spherical harmonic index $l$, which significantly slows down the computation.

\section*{Acknowledgements}

The work of OL is  partially financed by the Ministry of Education and Science of the Republic of Kazakhstan, Grant: IRN AP19680128. S.M. acknowledges financial support from "PNRR MUR project PE0000023-NQSTI".

\appendix

\section{Estimating the length at which quantum effects should dominate}\label{sec:RstarEstimation}

Starting from the Heisenberg uncertainty principle, we write
\begin{equation}\label{eq:HUP}
    \sigma_x\sigma_p\geq\frac{\hbar}{2},
\end{equation}
where sigma denotes the standard deviation of the observable reported in the subscript. Note that in this section we are using SI units, in order to emphasise the intertwining between gravity ($G$) and quantum mechanics ($\hbar$).

This relation implies that we cannot confine a particle in an arbitrarily small volume, since its momentum would become arbitrarily high. In standard quantum mechanics, the principle is encoded in the commutation properties of the observables. We now would like to obtain a potential-based description of this principle. Consider a free particle confined in a sphere of radius $r$. We can approximate its position uncertainty with the size of the sphere, hence $\sigma_x \propto r$. We also assume\footnote{This is justified a posteriori by the fact that the formula gives the correct result for the hydrogen $1s$ orbital radius.}  $\sigma_p\propto p$, where $p$ is the particle momentum. We can then rewrite the uncertainty principle as $pr\geq \hbar$. This approach has already been successfully used to intuitively explain the Hawking radiation and the Unruh effect \cite{Scardigli:1995qd}. We now want to consider a Schwarzschild black hole. If matter survives in any form after entering the event horizon, it should be made of some kind of particles which carry a mass $m$ and have a position $r$. We then assume that near the origin, the only effect that prevents the collapse of all the matter into a point is the uncertainty principle. Let also assume that we are in the situation with minimum uncertainty, then Eq. (\ref{eq:HUP}), under the considered hypothesis, leads to a potential
\begin{equation}\label{eq:HeisenbergPotential}
\begin{aligned}
     \phi_H(r)&=\frac{p^2}{2m}\propto\frac{\hbar^2}{mr^2},\quad \text{for massive particles,}\\
     &=pc\propto\frac{\hbar c}{r},\quad \text{for photons.}\\
\end{aligned}
\end{equation}
The effective repulsive potential due to uncertainty, denoted as $\phi_H$ where $H$ stands for Heisenberg, is proportional to $S^{-1}$ for massive particles, where $S$ is the area of the spherical region. A simple sanity check involves equating $\phi_H$ to the electric energy between a proton and an electron, and solve for the radius. The result indicates that the electron \virgolette{hovers} at a Bohr radius, $a_e$, from the nucleus. Since for electric forces we obtain the correct result, to estimate the radius at which quantum effects surpass gravitational collapse we set the Heisenberg energy equal to the gravitational energy
\begin{equation}\label{eq:gravBohrRad}
a_g = \frac{\hbar^2}{GMm^2} \propto l_{\mathrm{pl}}\left(\frac{l_{\mathrm{pl}}}{r_s}\right)\left(\frac{m_{\mathrm{pl}}}{m}\right)^2.
\end{equation}
Here, $a_g$ is analogous to the Bohr radius, which approximates the size of the $1s$ orbital for electric interactions. Since we assumed a zero angular momentum for the particle, $a_g$ should in turn approximate the size of the $1s$ gravitational orbital. This analogy could be further developed by defining a gravitational fine structure constant \cite{pellis2022unity,terazawa1977simple,Jentschura:2014uwa}, but that is outside the scope of this paper.

We now need to address the issue of the absence of evidence for a fundamental mass, since this prevents us from uniquely defining $a_g$. Since there is no convincing literature on either the value or the existence of a fundamental mass analogue to the fundamental charge, we opt for the electron mass as the fundamental mass scale, given that it is the smallest known particle mass. With this choice we are most probably overestimating the value of $m$ since if the fundamental mass had a value close to the electron's mass, it would be experimentally detected. We want to clarify that the exact value of $m$ in our equation (\ref{eq:heisenberg}), does not fundamentally affect the the entanglement entropy. This is because, in all considered metrics, the field-curvature coupling constant $\xi$ and $m$ are degenerate parameters. We report the dependence of the field-curvature coupling term $\xi R(r)$, close to the origin, on $m$ and $\xi$:
\begin{itemize}
    \item[-] Quantum-corrected Schwarzschild:  $ \xi/m^6$,
    \item[-] Coherent quantum: $\xi m^2$,
    \item[-] Hayward: $\xi m^4$,
    \item[-] corrected-Hayward: $\xi m^8$.
\end{itemize}

\subsection{Other arguments to compute the quantum length scale}

In this section, we reintroduce Planck units for simplicity. We aim to investigate alternative methodologies for determining the distance from the origin at which quantum effects surpass classical general relativity in quantum-corrected Schwarzschild black holes.

\subsubsection{Trace anomaly}

In Ref. \cite{Abedi:2015yga}, the impact of the trace anomaly \cite{Mottola:2010gp,PhysRevD.63.083504,PhysRevD.15.2088} on the collapse of a black hole is investigated. Their findings indicate that the metric is modified by the vacuum energy of quantum fields, resulting in a repulsive potential barrier. Consequently, a falling mass $m$ halts its motion before reaching the origin, at a radius $a_g \approx \left(c_A M\right)^{1/3}$. When considering a single infalling spin-zero particle, for which $c_A=1$, we find that $a_g \approx \left(\frac{r_s}{2}\right)^{1/3}$. In this case, we do not have the dependence of $a_g$ on the particle mass.

\subsubsection{Graviton corpuscolar scaling}

Another strategy to estimate $a_g$ is given in Ref. \cite{casadio2022geometry}: in order to have the corpuscular scaling \cite{Bose:2021ytn,PhysRevD.97.044047} for the number of gravitons is required that $R_S\approx r_s/\ln(R_\infty/r_s)$, where $r_s$ is the black hole horizon radius. This condition implies a dependence of the quantum matter core size\footnote{See Appendix \ref{sec:cohe_quant}.} on the radius of the outer region. Setting the infrared cutoff, $R_\infty$, equal to the observable universe radius yields $a_g=r_s/60$.

\subsubsection{Asymptotic safe quantum gravity}

Yet another perspective is given by Ref. \cite{Pawlowski:2018swz}. They work in the context of asymptotically safe quantum gravity. Using their form for the running gravitational constant and parameters values, we can estimate $a_g\approx r_s^{\frac{1}{3}}$.

Summarising we have that
\begin{align}
    a_g &\approx \frac{1}{r_s m^2} \quad \text{(Heisenberg uncertainty)} \label{eq:heisenberg} \\
    a_g &\propto r_s^{\frac{1}{3}} \,\,\qquad \text{(Trace anomaly \& asymptotic safety)} \label{eq:trace} \\
    a_g &\approx \frac{r_s}{60} \,\qquad \text{(Corpuscular scaling constraint)} \label{eq:corpuscular}
\end{align}

The take-home message is that quantum effects are still relevant at lengths much higher than the Planck scale for black holes with $r_s\gg 1$. It is also interesting to note that the trace anomaly and asymptotic safety approach yield the same functional dependence of $a_g$ on $r_s$.

\section{Numerical considerations}\label{sec:num_consid}

When an inner horizon is present or when the scalar curvature blows up near the origin, we cannot directly apply Eq. (\ref{kmat}). In those cases, we must consider $i,j=n_{\mathrm{min}},\hdots,n_{\mathrm{min}}+N=n_{\mathrm{max}}$ instead of $i,j=1,\hdots,N$, where $N$ is the total number of shells. In this way we consider the harmonic oscillators within a region free from pathological behaviours which can result in physical and numerical errors.

\subsection{Sum truncation}

\begin{figure}[H]
    \centering
    \includegraphics[width=1\columnwidth,clip]{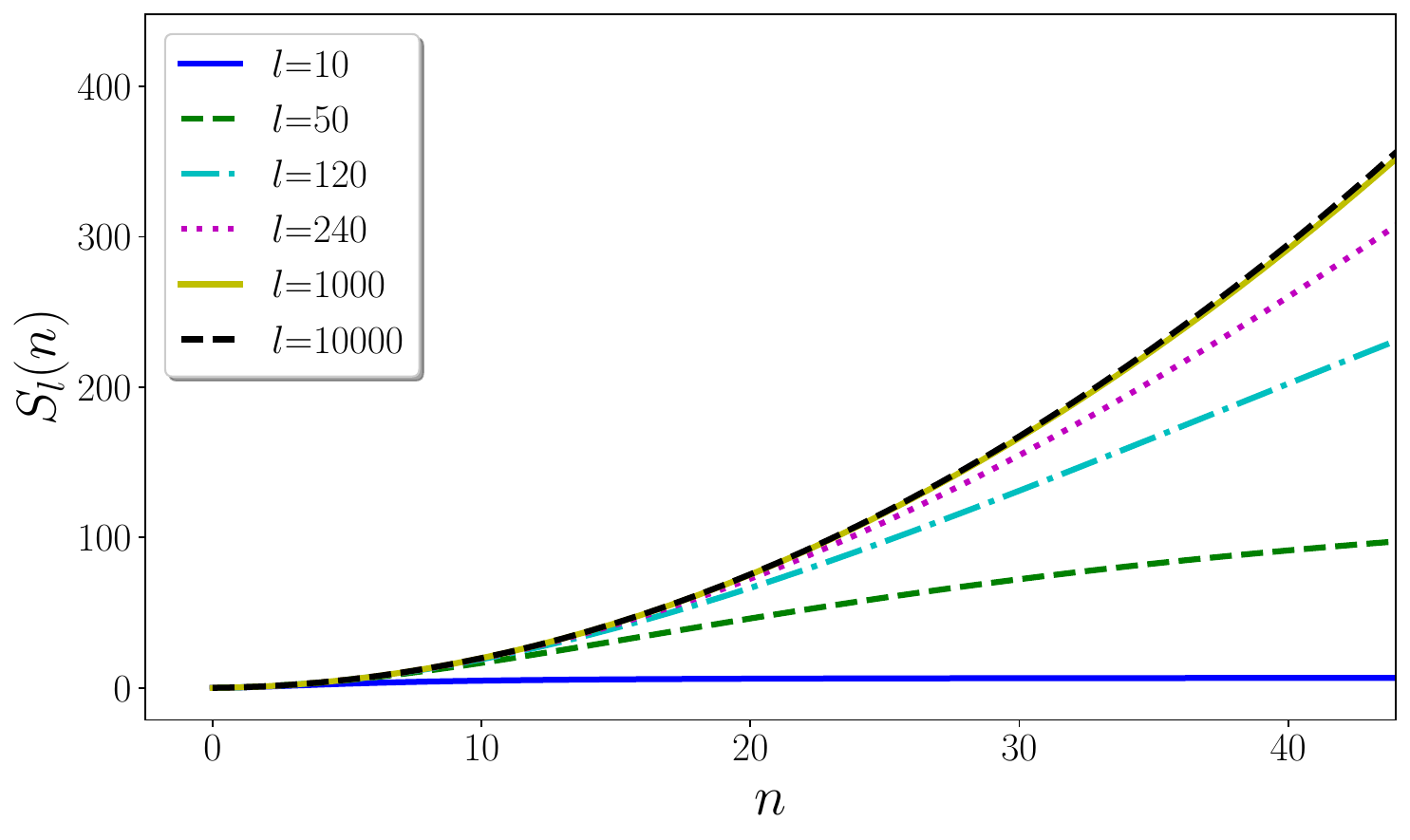}
    \includegraphics[width=1\columnwidth,clip]{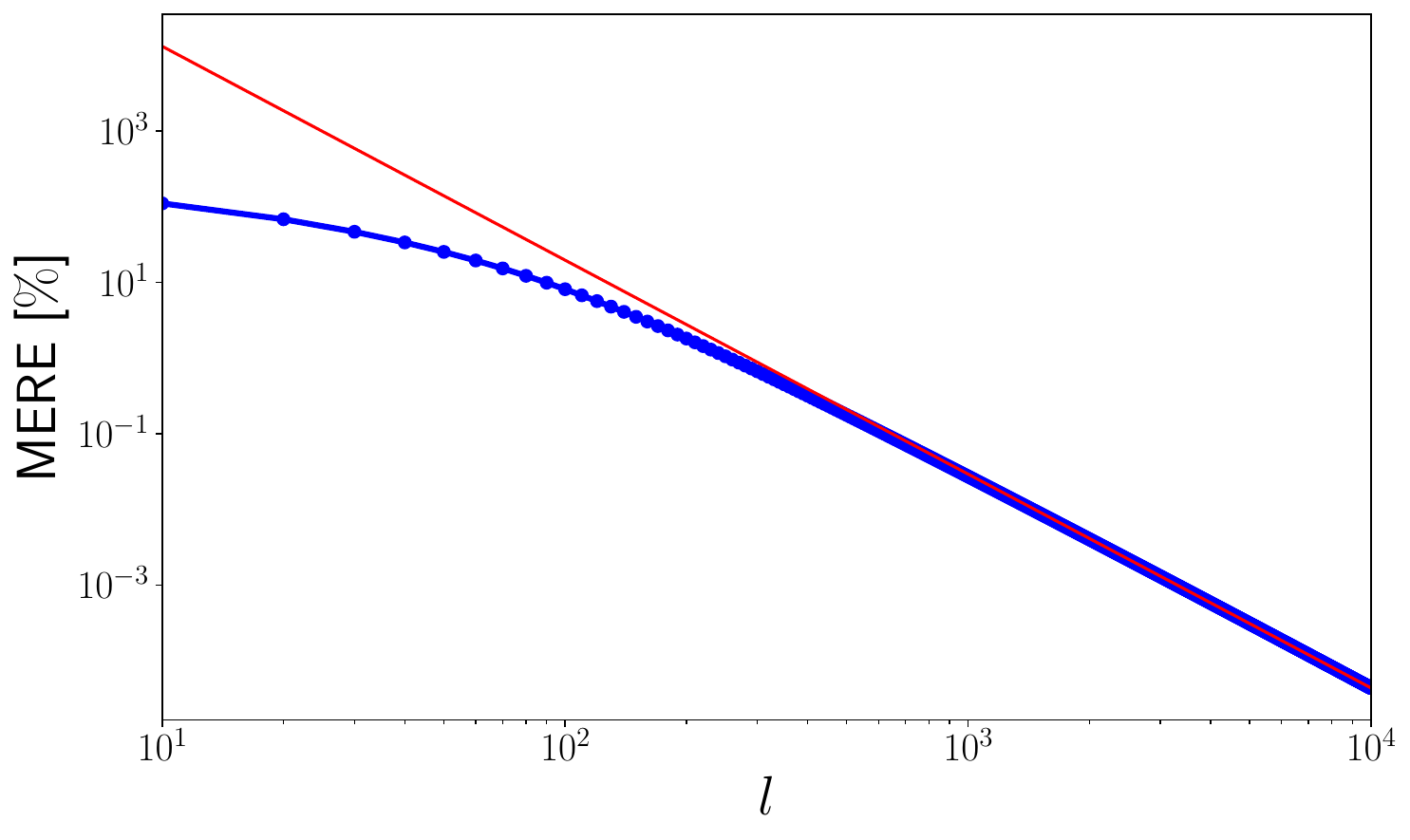}
    \caption{\textbf{\textit{Top}} : Entanglement entropy in flat spacetime for different values of the sum truncation parameter. \newline \textbf{\textit{Bottom}} : Entanglement entropy maximum estimated relative error (MERE) in flat spacetime. The red solid line is the linear fit.
    The parameters used in both plots are $a=1$, $m=1$, $n_{\mathrm{min}}=1$, $N=50$, $\Delta l=10$. A tolerance of $0.1\%$ is achieved with $l_\mathrm{max}=610$. Note that with these parameters, in log-log scale the MERE tends to become linear for large $l$.}
    \label{fig:lmax}
\end{figure}

\begin{figure}[H]
    \centering
    \includegraphics[width=1\columnwidth,clip]{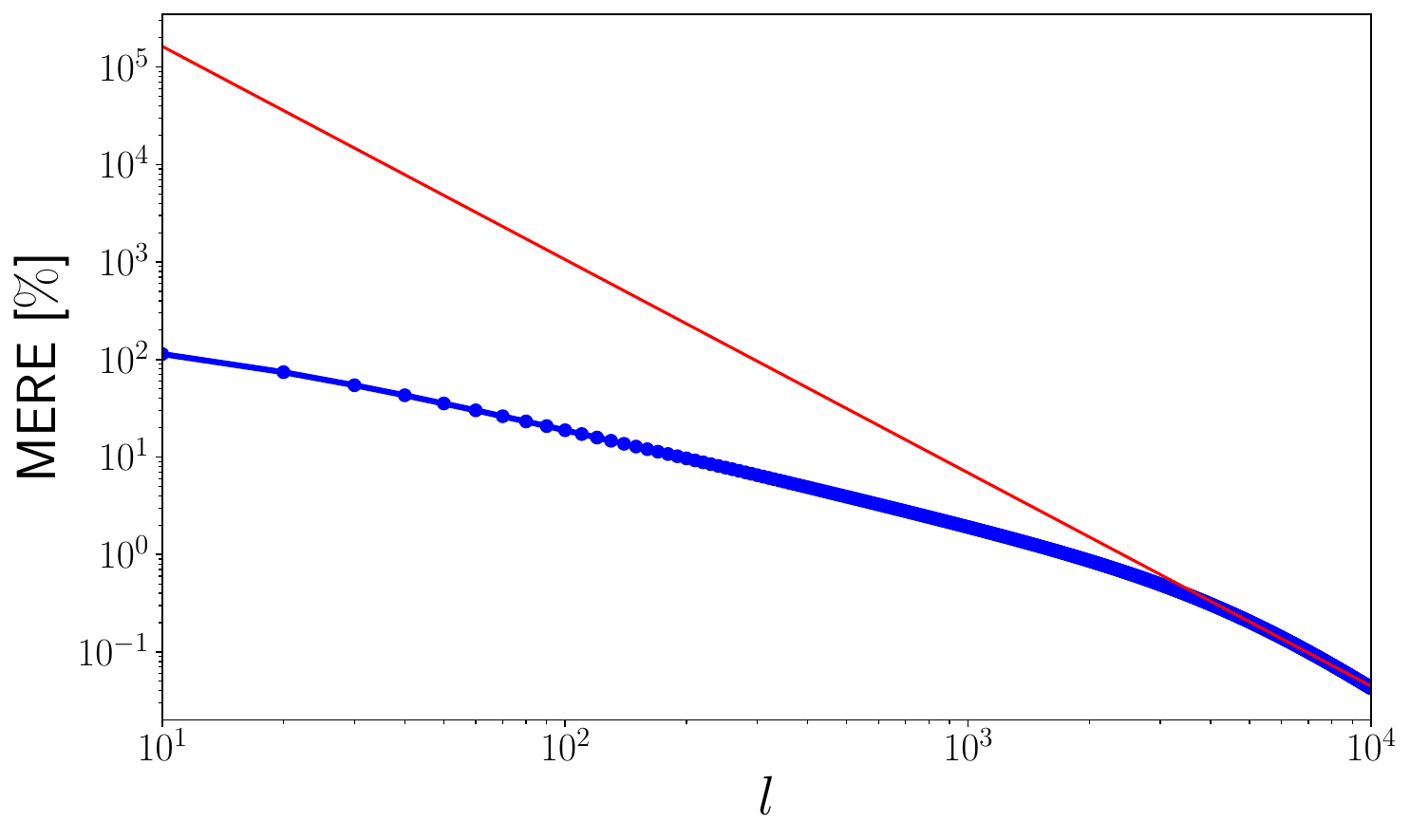}
    \caption{Entanglement entropy maximum estimated relative error (MERE) in flat spacetime. The red solid line is the linear fit. The used parameters are $a=1$, $m=10^2 $, $n_{\mathrm{min}}=0$, $N=50$, $\Delta l=10$.  A tolerance of $0.1\%$ is achieved with $l_\mathrm{max}=7010$. Note that the linear behaviour of Fig. \ref{fig:lmax} is still present but it seems to begin at higher $l$.}
    \label{fig:lmax_large_maxx}
\end{figure}

An important parameter to consider when numerically computing the entanglement entropy is the sum truncation value. Since we cannot sum the infinite terms appearing in Eq. (\ref{eq:ent_entropy_nl}), we must appropriately choose an $l_{\mathrm{max}}$ in such a way to obtain the desired tolerance. This can be easily done by computing the following function
\begin{equation}\label{eq:MaxRelErr}
    {\rm max}_{n \in [0,N]}\left|\frac{S_{l+\Delta l}(n)-S_{l}(n)}{\frac{S_{l+\Delta l}(n)+S_{l}(n)}{2}}\right|,
\end{equation}
where $l$ is the sum truncation value, which now is the independent variable, $n$ is the number of traced out shells among the $N$ available, and $\Delta l$ is the step size. The meaning of Eq. (\ref{eq:MaxRelErr}) is to give a percentage, which quantifies the difference between $S_{l+\Delta l}(n)$ and $S_{l}(n)$, with respect to their average. If the method converges, the numerator, and thus Eq. (\ref{eq:MaxRelErr}), approaches zero. We decided to take the maximum over $n$ to ensure that we are not underestimating the error. Results are shown in Fig. \ref{fig:lmax}.
Another detail one should be aware of is that the $l_\mathrm{max}$ computed in Fig. \ref{fig:lmax}, depends on the field mass. As an example we show in Fig. \ref{fig:lmax_large_maxx}, that switching from $m=1$ to $m=10^2$ and maintaining the same tolerance, $0.1\%$, we need to switch from $l_{\mathrm{max}}=610$ to $l_{\mathrm{max}}=7010$.

\subsection{Area law and field mass}\label{sec:num_consid_mass}

\begin{figure}[H]
    \centering
    \includegraphics[width=1\columnwidth,clip]{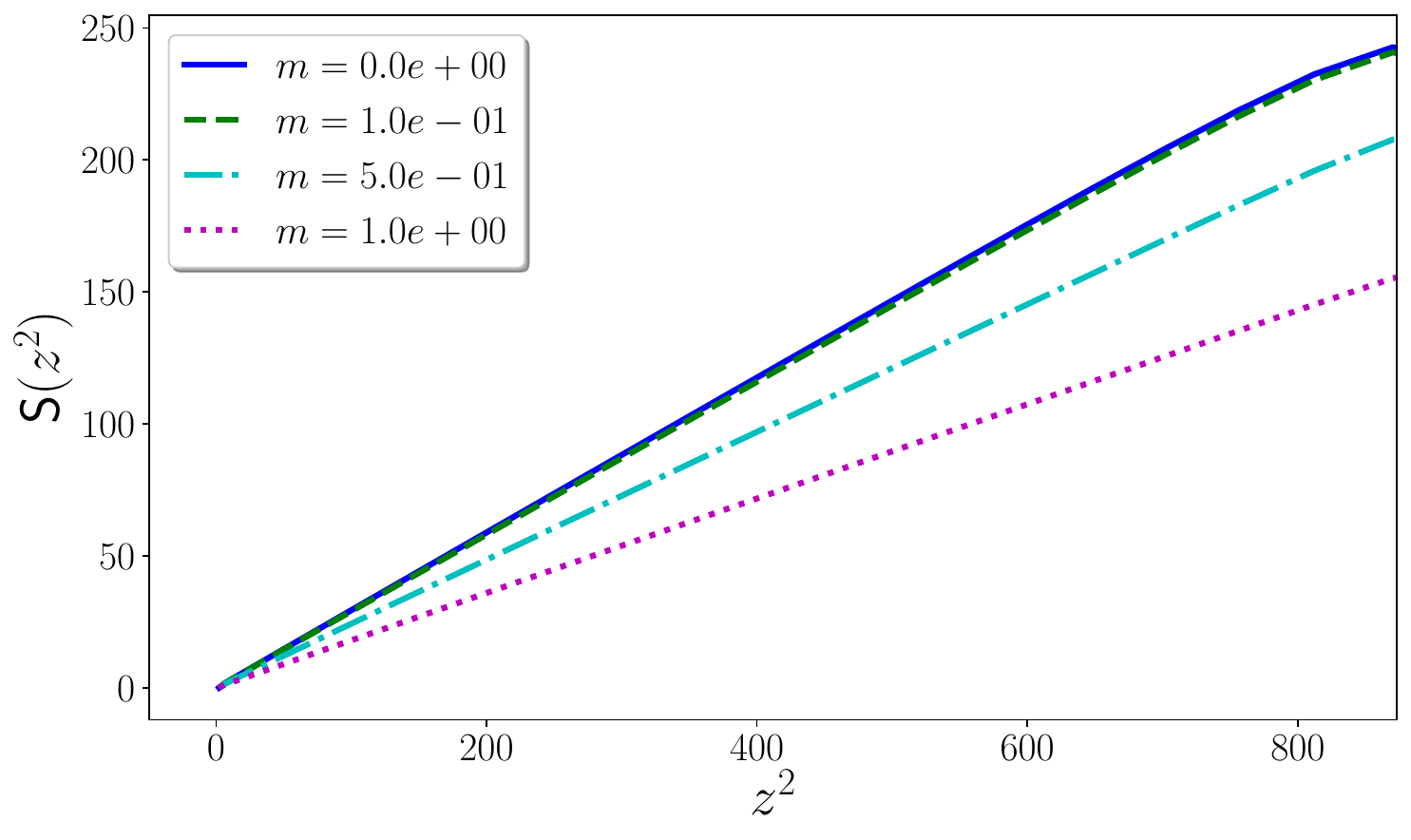}
    \caption{Entanglement entropy in Minkowski spacetime for different values of the field mass. The parameters used are $a=1$, $n_{\mathrm{min}}=1$, $N=30$, $l_\mathrm{max}=10^3$.}
    \label{fig:flat_mass}
\end{figure}

As shown in Fig. \ref{fig:flat_mass}, the entanglement entropy begins to significantly change slope for $m\gtrsim 1$.
We therefore made sure that the effective mass covers a suitable range, considering both $m_{\mathrm{eff}}\gg 1$ and $m_{\mathrm{eff}}\ll 1$, in order to clearly show the effect of the field-curvature coupling. Furthermore, $m_{\mathrm{eff}}\approx 1$ can be taken as the \virgolette{mid-way} value that separates the massless from the massive behaviours.

\subsection{Advantages of scale invariant curvature}

For a function $f$ to be scale invariant means that $f(\lambda x)=\lambda^\Delta f(x)$, for some fixed exponent $\Delta$ and all scaling factor $\lambda$. If the curvature is scale invariant, then the effective mass transforms under scaling as $m_{\mathrm{eff}}^2(\lambda r)=m^2+\xi\lambda^{\Delta} R(r)$. This property is quite nice since the scale of the black hole degenerates with the coupling constant. In fact, if we consider a black hole $\lambda$ times bigger(smaller), we can \virgolette{renormalize} the coupling, $\xi\rightarrow\xi/\lambda^{\Delta}$, so that the effective mass remains unchanged. Among the metrics considered in this work, all of them except the Hayward metric are scale invariant near the origin.

\subsection{Parameter values choices}\label{sec:num_cons_params}

To generate the plot depicted in Fig. \ref{fig:eeQ}, we opted for a small black hole since in Ref. \cite{Binetti:2022xdi} works with black holes characterised by few Planck masses. In order to show that the area law is restored far away from the origin, we require that $\xi R(N)\ll 1$, which sets an upper bound for the coupling strength $\xi$. Since we are considering a metric that reduce to Schwarzschild, we have that $\lim_{N\rightarrow\infty}R(N)=0$. The coupling upper bound would be infinite, but since we are limited to $N\approx 100$, we have to set a small coupling constant if we want to show that the area law is eventually restored and the coupling constant should be smaller for bigger black holes.

\section{Derivation of coherent quantum black hole metric}\label{sec:cohe_quant}

Here we summarize how Eq. (\ref{eq:cohLapse}) is derived, as discussed in Ref. \cite{casadio2022geometry}. We first show how to describe a generic static, spherically symmetric potential, $V(r)$, as the mean field of the coherent state of a free massless scalar field.

We rescale the potential by defining the field $\phi=\sqrt{m_{\mathrm{pl}}/l_{\mathrm{pl}}}V$. The scalar field is assumed to satisfy the massless Klein-Gordon equation
\begin{equation}\label{eq:masslessKGSph}
    \left[\partial_t^2-\frac{1}{r^2}\partial_r(r^2\partial_r)\right] \phi =0.
\end{equation}
In spherical coordinates, solutions to Eq. \eqref{eq:masslessKGSph} are given by
\begin{equation}
    u_k(t,r)=e^{-ikt}j_0(kr),
\end{equation}
where $k>0$ and $j_0=\sin(kr)/kr$ is a spherical Bessel function. We can therefore quantise the field by writing down the mode expansion of the field operator
\begin{equation} \label{scalexpa}
    \hat{\phi}=\int_0^\infty \frac{k^2 \mathrm{d}k}{2\pi^2}\sqrt{\frac{\hbar}{2k}}\left[\hat{a}_k u_k(t,r)+\hat{a}^\dag_k u^*_k(t,r)\right].
\end{equation}
The Fock space of quantum states is built as usual from the vacuum state, defined by $\hat{a}_k\ket{0}=0$, for all $k>0$. We have assumed Minkowski spacetime in Eq. (\ref{eq:masslessKGSph}) since this vacuum choice should describe a spacetime devoid of any energy.

A natural choice to describe classical configurations of the scalar field are coherent states. In fact, coherent states posses the \virgolette{most classical} behaviour. For example, coherent states have minimum uncertainty and the time evolution of a coherent state is concentrated along the classical trajectories.
\begin{equation}
    \hat{a}\ket{g}=g_ke^{i\gamma_k(t)}\ket{g},
\end{equation}
where $g_k$ and $\gamma_k(t)$ are free functions we will shortly determine.
We thus seek the coherent state giving rise to the potential $V(r)$, namely
\begin{equation}
    V(r)=\sqrt{\frac{l_{\mathrm{pl}}}{m_{\mathrm{pl}}}}\bra{g}\hat{\phi}\ket{g}.
\end{equation}
Substituting the field expansion of Eq. \eqref{scalexpa}, we obtain
\begin{equation}\label{eq:quantPotent}
    V(r)=\int_0^\infty \frac{k^2 \mathrm{d}k}{2\pi^2}l_\mathrm{pl}\sqrt{\frac{2}{k}}g_k\cos[\gamma_k(t)-kt]j_0(kr).
\end{equation}
If we now define
\begin{equation}
    V(r)=\int_0^\infty \frac{k^2 \mathrm{d}k}{2\pi^2}l_\mathrm{pl}\sqrt{\frac{2}{k}}\Tilde{V}(k)j_0(kr)
\end{equation}
we obtain the constrains $\gamma_k(t)=kt$ and $g_k=\sqrt{\frac{k}{2}}\frac{\Tilde{V}(k)}{l_\mathrm{pl}}$. However, we cannot impose that the average of the field yields exactly the gravitational potential, as this would result in unacceptable $k$-mode occupation numbers $g_k$. Since we are interested in reproducing the black hole gravitational potential only outside the event horizon, we impose the less restrictive condition on the coherent state $\ket{g_{BH}}$ representing the black hole
\begin{equation}
    \sqrt{\frac{l_{\mathrm{pl}}}{m_\mathrm{pl}}}\bra{g_{BH}}\hat{\phi}(t,r)\ket{g_{BH}}\approx V_{N}(r) \quad \text{for} \quad r>R_H,
\end{equation}
where $V_N$ is the classical Newtonian potential. Using this relaxed condition, the state can be properly normalised. In fact, this condition means that the state $\ket{g_{BH}}$ does not contain modes of arbitrarily small wavelengths needed in order to resolve the classical singularity. We therefore introduce the ultraviolet cutoff $R_S$, so that Eq. (\ref{eq:quantPotent}) yields
\begin{equation}
    V_{QN}\approx -\frac{2GM}{\pi r}\text{Si}\left(\frac{r}{R_S}\right),
\end{equation}
where Si is the sine integral function. The quantity $R_S$ physically represent the size of the quantum matter core. Such potential allows us to construct the coherent black hole metric lapse function as $f(r)=1+2V_{QN}(r)$.


\section{Bound on power law corrections to the Schwarzschild metric}\label{sec:bound}

We present a brief review of the main results of Ref. \cite{Knorr:2022kqp}, where regular black hole solutions are investigated assuming the existence of an action principle in quantum gravity. In particular, we are interested in the asymptotic behaviour ($r\rightarrow \infty$) of a spherically symmetric, static background.  We then assume that the metric has a quantum correction in the form of a power-law with exponent $n$
\begin{equation}\label{eq:mod_schw}
    f(r)=1-\frac{r_s}{r}+\frac{c}{r^n}.
\end{equation}
We impose that this metric can be obtained as a solution of the effective field equations stemming from a quantum gravity effective action.
The ansatz for the effective action is
\begin{equation}\label{eq:effective_action}
\begin{aligned}
        \Gamma &= \frac{1}{16 \pi}\int \mathrm{d}^4x \sqrt{-g}\big[-R
        \\& -\frac{1}{6}Rf_R(\Delta)R+R^\mu_\nu f_{Ric}R^\mu_\nu +\mathcal{O}(R^3)\big]
\end{aligned}
\end{equation}
where $\Delta=-g^{\mu\nu}D_\mu D_\nu$ is the d'Alembert operator of the metric $g$ and $R$ is a generic curvature tensor. The functions $f_{R}$ and $f_{Ric}$ are known as form factors.

The equation of motion for the metric is given by
\begin{equation}
    \frac{\delta \Gamma(g)}{\delta g_{\mu\nu}}=0.
\end{equation}
We approximate both the form factors with a truncated Taylor expansion. Since the action is a sum of terms, also the variation of the action will be a sum of terms of the form
\begin{equation}
    \delta \Gamma(g)=\delta \Gamma_{GR}(g)+\delta \Gamma_{R^2}(g)+\delta \Gamma_{R^3}(g)+\hdots
\end{equation}
The first term simply arises from general relativity. To leading order, at large $r$ we find
\begin{equation}
    \frac{\delta \Gamma_{GR}(g)}{\delta g_{00}}\Bigg |_{GR}\propto \frac{\delta \Gamma_{GR}(g)}{\delta g_{11}}\Bigg |_{GR}\propto r^{-n-2}.
\end{equation}
Thus Eq. (\ref{eq:mod_schw}) can be a solution if the other terms in the effective action can cancel those general relativity terms. However, it can be shown that a truncated Taylor expansion for the form factors is not able to remove such GR terms. The only contributions that are capable of removing the GR terms are those constructed from the Weyl tensor and at most one occurrence of the Ricci scalar. The lowest order correction for which such a cancellation is possible reads
\begin{equation}
\begin{aligned}
        \Gamma^{loc}_{R^3}&=\frac{1}{16 \pi}\int \mathrm{d}^4 x\sqrt{-g}\big[ c_1 R C^{\mu\nu\rho\sigma}C_{\mu\nu\rho\sigma}+\\
        & +c_2 C_{\mu\nu}^{\rho\sigma}C_{\rho\sigma}^{\tau\omega}C_{\tau\omega}^{\mu\nu}\big]
\end{aligned}
\end{equation}
This term gives a contribution to the equation of motion $\propto r^{-8}$. Thus it can cancel the $GR$ terms if $n=6$. Using higher order terms we can cancel other specific powers $n$. In fact it can be shown that cancellation is possible for all $n\geq 6$, $n\neq 7$. The Hayward metric do not satisfy this bound, since in this case the correction is proportional to $r^{-4}$. A modification of the Hayward metric which instead satisfy the bound is
\begin{equation}
    f(r)=1-\frac{r_s r^4}{r^5+r_s b^4},
\end{equation}
which has been considered in Sec. \ref{Sezdisc} to compute the scalar field entanglement entropy.

\bibliographystyle{unsrt}

\end{document}